\documentclass[pdflatex,sn-mathphys-ay,iicol]{sn-jnl}%

\usepackage{graphicx}%
\usepackage{multirow}%
\usepackage{amsmath,amssymb,amsfonts}%
\usepackage{amsthm}%
\usepackage{mathrsfs}%
\usepackage[title]{appendix}%
\usepackage{xcolor}%
\usepackage{textcomp}%
\usepackage{manyfoot}%
\usepackage{booktabs}%
\usepackage{algorithm}%
\usepackage{algorithmicx}%
\usepackage{algpseudocode}%
\usepackage{listings}%

\usepackage{natbib}
\usepackage[utf8]{inputenc}
\usepackage{amsmath,amsthm,mathtools}
\usepackage{subcaption}
\usepackage{graphicx,tikz,float}
\usetikzlibrary{arrows,positioning,calc}

\usepackage{amssymb,color}

\usepackage{enumitem}

\usepackage[english]{babel}
\usepackage{graphicx,epstopdf,float,tikz}
\usetikzlibrary{arrows,positioning,calc}

 \newenvironment{syseq*}{%
 	\color{gray} \left\{ \normalcolor
 	\begin{aligned}
 	}{%
 	\end{aligned} \right. }
 
\newtheoremstyle{mytheoremstyle} 
{\smallskipamount}                    
{\smallskipamount}                    
{\itshape}                   
{}                           
{\bfseries}                   
{.}                          
{.5em}                       
{}  
 
\theoremstyle{mytheoremstyle}
 
\newtheorem{proposition}{Proposition} 

\newtheorem{theorem}{Theorem}

\newtheorem{corollary}{Corollary}

\newcommand{\R}{\mathbb R}

\newcommand{\N}{\mathbb N}

\newcommand{\cS}{\mathcal S}

\newcommand{\x}{\times}
\newcommand{\st}{\mid} 
\newcommand{\inv}{^{-1}}

\DeclareMathOperator{\diag}{diag}

\newcommand{\emquote}[1]{\emph{``#1''}}

\newcommand{\sr}{^\star}  
\newcommand{\T}{^\top}  

\newcommand{\e}{\text{\tiny E}}
\renewcommand{\i}{\text{\tiny I}} 
\renewcommand{\u}{\text{\tiny U}}

\definecolor{dark-magenta}{RGB}{176, 70, 161}

\allowdisplaybreaks

\graphicspath{{figs/}}

\begin{document}

\title[A state-space characterization of reliability in FitzHugh-Nagumo neurons and networks via contraction analysis]{A state-space characterization of reliability in FitzHugh-Nagumo neurons and networks via contraction analysis}


\author[1]{\fnm{Michelangelo} \sur{Bin}}

\author[1,2]{\fnm{Alessandro} \sur{Cecconi}}

\author[1]{\fnm{Lorenzo} \sur{Marconi}}

\affil[1]{\orgdiv{Department} of Electrical, Electronic and Information Engineering, \orgname{University of Bologna}, \city{Bologna}, \country{Italy}}

\affil[2]{\orgdiv{Department} of Electrical Engineering, \orgname{KU Leuven}, \city{Leuven}, \country{Belgium}}


\abstract{A new contraction-theoretic state-space characterization of neuronal reliability is developed for FitzHugh-Nagumo models. It is shown that the state space of a neuron can be partitioned into two contraction regions separated by an expansion region where trajectory expansion may occur. The unforced resting equilibrium lies in a contraction region, and each full spike drives the state to the other contraction region and then back to the first. In-between, the state crosses the expansion region. Over any given interval, reliability is characterized as an average contraction property given by the balance  of the contraction and expansion accumulated along the trajectories. An exact integral characterization of this balance and a simpler sufficient dwell-time condition are derived. The framework is then extended to networks of FitzHugh-Nagumo systems with an arbitrary affine voltage-coupling topology. A network of $n$ neurons is shown to possess $2^n$ contraction regions, separated by an expansion region, and network reliability is again characterized in terms of average contraction. A proof-of-concept excitatory-inhibitory network example is used to illustrate a connection between this characterization of reliability and regulation in neuronal networks. The network uses an internal model of a class of inputs to create a homeostatic regime where the network output is insensitive to the specific input within that class. Reliability is then seen as providing the  asymptotic convergence properties necessary to make such a target regime attractive. Simulations of multi-stage cascade compositions of this elementary homeostatic motif illustrate how regulation can support oscillatory-signal propagation without phase-lag accumulation.}




\maketitle

\section{Introduction}\label{sec.intro}


Reliability -- the reproducibility of a neuron's spike train across repeated presentations of the same input -- is a widely studied phenomenon in neuroscience. Reliable spike timing has been observed experimentally under several stimulation conditions \citep{mainen_reliability_1995,hunter_resonance_1998}, and the experimental evidence is complemented by numerous theoretical and numerical studies \citep{kosmidis_analysis_2003,ritt_evaluation_2003,brette_reliability_2003,zhou_noise-induced_2003,goldobin_antireliability_2006,ermentrout_reliability_2008,sun_pseudo-lyapunov_2010,lin_stimulus-response_2013}.
In these studies,  reliability is typically formalized in terms of the loss of sensitivity to initial conditions under common forcing, and the formal arguments are based on specific neuron models, including the \emph{FitzHugh-Nagumo} system \citep{kosmidis_analysis_2003,goldobin_antireliability_2006} also considered  here, simpler one-dimensional phase models \citep{ritt_evaluation_2003}, and the more detailed \emph{Hodgkin-Huxley} model \citep{zhou_noise-induced_2003,sun_pseudo-lyapunov_2010}. 

A common analysis approach to characterize reliability is  through the maximal \emph{Lyapunov exponent} of the response dynamics, which measures the asymptotic average exponential rate at which infinitesimal perturbations of a response trajectory grow or decay.
A negative maximal Lyapunov exponent therefore provides evidence of local convergence under common forcing.
However, Lyapunov exponents have several limitations in terms of mathematical guarantees and computational aspects, among which: \emph{(i)}  negativity of the maximal Lyapunov exponent along a given trajectory does not always guarantee  stability of the trajectory due to the \emph{Perron effect} \citep{leonov_time-varying_2007};  \emph{(ii)} they are ``average'' quantities summarizing the expansion and contraction phases over time, and they thus do not identify when these two behaviors occur along the trajectory; \emph{(iii)} except for a few analytically tractable cases, they must be estimated numerically, typically through long-horizon integration of the state and its variational dynamics.

In addition, many existing approaches assume a common  ``frozen-noise'' input sampled from a prescribed stochastic process. Their conclusions are therefore generally tied to the assumed input statistics and  do not, by themselves, provide characterizations that hold for arbitrary, possibly deterministic inputs.


In the simplifying, yet emblematic  context of FitzHugh-Nagumo (FHN) models,  this paper  develops  a new state-space characterization of  reliability   based on \emph{contraction analysis}~\citep{lohmiller_contraction_1998,bullo_contraction_2024}. 
It is shown that the state space of a single FHN system is partitioned into three regions: two ``contraction regions'', in which trajectories converge toward one another, separated by an ``expansion region'', in which they may diverge. 
The unforced system's resting point lies in one of the two contraction regions, and each spike brings the system's state to the other and then back to the first. In-between, the state necessarily crosses the expansion region. Over any given time interval, inputs that keep the considered trajectories together within the same contraction region for a sufficiently large fraction of time induce a net contraction over that interval. This ``average contraction'' is used to characterize a reliable behavior, as it implies a progressive loss of dependence on the initial conditions. 

This framework is then extended to a class of networks formed by an arbitrary number of FHN models interconnected through affine couplings in the voltage-like variables organized according to an arbitrary topology. The same geometric characterization of the state space in terms of contraction and expansion regions is retained in the network case. In particular, a network of $n$ neurons exhibits $2^n$ contraction regions separated by an expansion region. Consequently, \emph{network reliability} is characterized in terms of average contraction  as in the single-neuron case.

This contraction viewpoint provides a geometric perspective whereby reliability is conceived in terms of specific regions of the state space, rather than solely through time- and trajectory-based indicators. In particular, it reveals that reliability emerges from an interplay of stability and instability: trajectory contraction takes place both near the resting point and during spiking.
Compared with approaches based on Lyapunov exponents, the proposed framework also provides sufficient conditions certifying reliability over finite time intervals and identifies when contraction and expansion phases occur along the trajectories.


Another interesting byproduct of contraction analysis is that it creates a link between reliability and regulation in neuronal networks.
In particular, it is well-known in the control theory literature~\citep{francis_internal_1975,pavlov_uniform_2006,pavlov_incremental_2008,bin_internal_2022,bin_about_2023,giaccagli_incremental_2024} that the design of a control system to solve a regulation problem -- namely to drive a system toward an asymptotic regime where the system behaves as desired -- can be accomplished with two ingredients. The first is a dynamical mechanism that is able to keep the controlled system in the desired working regime. This typically employs an \emph{internal model} of the exogenous disturbances that must be counteracted and of the desired motions that must be performed~\citep{francis_internal_1975,bin_internal_2022}. The second ingredient is a \emph{stabilization} mechanism ensuring that the target asymptotic regime is robustly reached. This can be ensured by inducing closed-loop contraction via feedback of the regulation error~\citep{francis_internal_1975,pavlov_uniform_2006,pavlov_incremental_2008,bin_about_2023,giaccagli_incremental_2024}, since contraction implies convergence of the trajectories to one another. As a consequence, in the presence of contraction, the existence of an ideal regime characterized by a desired functionality -- a consequence of the presence of a proper internal model -- becomes a sufficient condition for its asymptotic achievement. 

This connection between reliability and regulation is investigated here  in the context of a proof-of-concept neuronal network example recently introduced in~\cite{cecconi2025NOLCOS}. This network consists of a basic excitatory-inhibitory motif made of three interconnected FHN systems that, when driven by a specific class of inputs, exhibits both homeostasis -- in the sense that a specific network variable converges to a steady state independent of the input -- and perfect synchronization with the input -- in the sense that another network variable  perfectly tracks the input with zero phase lag.
Using the proposed contraction-based characterization of reliability, it is   shown that, for a class of input stimuli, reliability over an infinite time horizon implies regulation toward the aforementioned homeostatic and synchronized regime. Owing to the control-theoretic analogy, this regulating behavior is connected to the internal model principle.

Finally, a series interconnection of several of these example networks is used to show that the same principles can lead to a network that asymptotically propagates a specified class of stimuli without accumulating phase lag across successive stages. This illustrates a new means of zero-lag propagation of oscillatory stimuli across neuronal networks.

The article is organized as follows: Section~\ref{sec.reliability} develops the contraction-based characterization of reliability for single FHN models. Specifically, Section~\ref{sec.reliability.contraction-notions} formalizes the employed contraction notions, Section~\ref{sec.reliability.contraction} develops the analysis of the FHN equations, and Section~\ref{sec.reliability.contraction-realiability} connects the proposed characterization to some experimental evidence.
Next, Section~\ref{sec.networks} extends the contraction-based characterization of reliability to the network case, while Section~\ref{sec.regulation} develops the previously-mentioned connection between reliability and regulation. Finally Section~\ref{sec.discussion} provides a concluding discussion.

The mathematical notation is standard. $\R$ and $\N$ denote the set of real and natural (including $0$) numbers, respectively. If $\sim$ is any relation on a set $S$, and $x\in S$, we let $S_{\sim x} \coloneqq\{ y\in S\st y \sim x\}$. Vectors of $\R^n$, where $n\in\N_{\ge 1}$, are denoted by $n$-tuples $(x_1,\ldots,x_n)$ with $x_i\in\R$ for each $i=1,\ldots,n$ and are equivalently treated as column vectors in matrix multiplication.  If $M_1,\cdots,M_n$ are matrices, $\diag(M_1,\ldots,M_n)$ denotes their block-diagonal concatenation. If $x\in\R^n$, $\diag(x)\coloneqq\diag(x_1,\ldots,x_n)$. For a symmetric matrix $M \in \R^{n\x n}$, $\lambda_{\max}(M)$ and $\lambda_{\min}(M)$ denote its largest and smallest eigenvalue, respectively. Given two symmetric matrices $A,B\in\R^{n\x n}$, $A\succ B$ ($A\succeq B$) means $x\T A x > x\T B x$ ($x\T A x \ge x\T B x$) for all $x\in\R^n\setminus\{0\}$.

\section{Contraction and Reliability} \label{sec.reliability}
 
\subsection{Contraction Notions} \label{sec.reliability.contraction-notions}
The notion of contraction used in this paper is an adaptation of the more standard notion used in~\cite{lohmiller_contraction_1998} and \cite{bullo_contraction_2024}.
A system $\dot x = f(x,t)$,
with state $x(t)\in\R^n$, is said to be \emph{contractive} on a convex set $C\subset\R^n$  with respect to a metric $d:\R^n \x \R^n\to[0,\infty)$  if there exists $\lambda>0$ such that, for every $t_0\in\R$, every two solutions $x_a$ and $x_b$ of the system satisfy $d(x_a(t),x_b(t))\le e^{-\lambda (t-t_0)}d(x_a(t_0),x_b(t_0))$
for every $t\ge t_0$ for which $x_a(\tau)\in C$ and $x_b(\tau)\in C$ for all $\tau\in[t_0,t]$. The set $C$ is called a \emph{contraction region} and the constant $\lambda$  the \emph{contraction rate}.
According to this definition, the distance between two trajectories decays exponentially on every time interval during which both trajectories remain in the same contraction region.

Furthermore, the following definitions provide a notion of contraction between trajectories over finite time intervals. Given two solutions $x_a$ and $x_b$, two time instants $t_0,t_1\in\R$ with $t_1>t_0$, and a scalar $\alpha\in[0,1)$, the solutions $x_a$ and $x_b$ are said to \emph{$\alpha$-contract from $t_0$ to $t_1$ with respect to $d$} if $d(x_a(t_1),x_b(t_1)) \le \alpha d(x_a(t_0),x_b(t_0))$. 
They are said to \emph{contract from $t_0$ to $t_1$ with respect to $d$} if $d(x_a(t_1),x_b(t_1)) < d(x_a(t_0),x_b(t_0))$.
Finally,  
$x_a$ and $x_b$ are said to \emph{asymptotically contract with respect to $d$} if $\lim_{t\to\infty} d(x_a(t),x_b(t)) = 0$.

In this article, these notions provide a sufficient state-space characterization of reliability. When the explicit time dependence of the system originates from the application of a common external input, trajectories starting from different initial conditions can be used to represent repeated time-aligned presentations of the same input. Asymptotic contraction then implies a progressive loss of dependence on the initial conditions and the asymptotic agreement of the response trajectories across trials.


 \subsection{Contraction of FitzHugh-Nagumo Models}\label{sec.reliability.contraction}

Consider the following FHN neuron model~\citep{fitzhugh_impulses_1961}
\begin{equation} \label{s.FN} 
	\begin{aligned}
	\dot v &= v-\tfrac{1}{3}v^3 -w + u, \\
	\dot w &= \varepsilon(v-b w+ a)  , 
	\end{aligned}
\end{equation}
with state $x=(v,w)\in\R^2$, and where
\begin{equation}\label{d.a_b_epsilon}
	1-2b/3 < a <1 ,\quad  b\in(0,1),\quad 0<\varepsilon<1/b.
\end{equation}    
The state $v$ represents a fast voltage-like activation variable,   $w$ represents a recovery variable that evolves on a slower time scale ($\varepsilon$ is typically small), and $u$ is the input stimulus.
Throughout the paper, inputs are assumed to be locally essentially bounded and the corresponding solutions are understood in the Carath\'{e}odory sense.  

For an arbitrary $\mu>0$, define the sets 
\begin{align*}
	C_{\downarrow}(\mu)  &\coloneqq \{ x\in\R^2 \st v\le -\sqrt{1+\mu}\} ,\\
	C^{\uparrow}(\mu)  &\coloneqq \{ x\in\R^2 \st v\ge  \sqrt{1+\mu}\}.
\end{align*} 
Both are convex and closed. 
Moreover, with $\varepsilon$ the same as in \eqref{s.FN}, define the following metric on $\R^2$ 
\begin{equation}\label{d.d}
	d(x_a,x_b) \coloneqq \sqrt{ \tfrac{1}{2} (v_a-v_b)^2 + \tfrac{1}{2\varepsilon}(w_a-w_b)^2}.
\end{equation}
Then, the following result  shows that, for every $\mu>0$, both  $C_\downarrow(\mu)$ and $C^\uparrow(\mu)$ are contraction regions for \eqref{s.FN}.
\begin{theorem}\label{thm.contraction}
	For every $\mu>0$, system~\eqref{s.FN} is contractive on both $C_\downarrow(\mu)$ and $C^\uparrow(\mu)$ with respect to the metric~\eqref{d.d} and with contraction rate $\lambda(\mu)\coloneqq\min\{\mu,b\varepsilon\}$.
\end{theorem}
 Theorem~\ref{thm.contraction} is proved in Appendix~\ref{sec.proof.thm.contraction}. It  implies that, for every $\mu>0$, every two solutions of \eqref{s.FN} converge  to one another in the metric \eqref{d.d} at exponential rate $\lambda(\mu)$   on any interval of time in which they lie in the same region $C_\downarrow(\mu)$ or $C^\uparrow(\mu)$. 
 Taking the union  for all $\mu>0$  leads to the definition of the following sets
 \begin{align*}
 	C_\downarrow &\coloneqq   \{ x\in\R^2\st v<-1\},&
 	C^\uparrow &\coloneqq   \{ x\in\R^2\st v>1\},
 \end{align*}  
 which we call the \emph{lower} and \emph{upper}  \emph{contraction region}, respectively.
 Clearly, if $x\in C^\uparrow$ (resp. $C_\downarrow$), then it is in $C^\uparrow(\mu)$ (resp. $C_\downarrow(\mu)$) for all small enough $\mu>0$. Hence, as long as two trajectories move together inside   $C^\uparrow$ or $C_\downarrow$, their distance decreases with time.
 However, it is important to notice that such a decrease may not happen at an exponential rate,  as the contraction rate vanishes if one of the two solutions approaches the boundary of $C^\uparrow\cup C_\downarrow$.  
 
 The complement of the union of the two contraction regions, namely $I \coloneqq \R^2 \setminus ( C^\uparrow\cup C_{\downarrow}    )	= \{x\in\R^2 \st v^2\le1\}$,
 defines a closed set where the FHN model \eqref{s.FN} is not contractive with respect to the metric \eqref{d.d}. This region is therefore called   the \emph{expansion region}, and it separates the two contraction regions. The terminology indicates that trajectories may diverge in $I$, although not every pair necessarily does so.

For $u=0$, the unforced system \eqref{s.FN} has an equilibrium $x\sr=(v\sr,w\sr)$ given by the real solution of $v^3 +\frac{3}{b}(1-b) v + 3\frac{a}{b} = 0$ and $v-bw+a=0$.
For the range of parameters \eqref{d.a_b_epsilon},   
one has 
$v\sr<-1$.  Indeed, by way of contradiction, assume that $v\sr\ge -1$. Then, the equilibrium conditions and \eqref{d.a_b_epsilon} imply $v\sr\in[-1,0]$ and, hence, $-1\le (v\sr)^3 = - \tfrac{3}{b}(1-b) v\sr - 3\frac{a}{b} \le \tfrac{3}{b}(1-b) -3\frac{a}{b}<-1$, a contradiction.
With standard nullcline arguments, it can also be shown that a full spike will bring $v$ to values larger than $1$.
Thus, the stable equilibrium $x\sr$ of the FHN model lies in the lower contraction region $C_{\downarrow}$ and every full spike brings the state to the upper contraction region $C^\uparrow$.   

A spike-inducing input has therefore the effect of making the system switch between the two contraction regions, from the lower to the upper, and back. In-between, the system crosses the expansion region $I$ where the trajectories may  diverge from one another. 
Over a time scale of multiple spikes, the overall contraction behavior depends on the balance between the contraction accumulated when the trajectories lie in the same contraction region and the possible expansion accumulated during the remaining time.
Specifically, pick an input $u$, and let $x_a$ and $x_b$ be any two solutions of \eqref{s.FN} corresponding to the input $u$. Let us associate with $x_a$ and $x_b$ the following functional 
\begin{equation}\label{d.nu}
	\begin{split}
		\nu(x_a,x_b)&\coloneqq  \frac{1}{3}(v_a^2+v_b^2+v_av_b-3)(v_a-v_b)^2 \\
		&\qquad+ b(w_a-w_b)^2.
	\end{split} 
\end{equation}
Pick two time instants $t_0,t_1\in\R$ such that $t_1\ge t_0$, and let $T_{\rm c}\subset[t_0,t_1]$ denote the set of time instants $t$ where $x_a(t)$ and $x_b(t)$ are both in $C^\uparrow$ or both in $C_\downarrow$. Finally, let $T_{\rm i}\coloneqq[t_0,t_1]\setminus T_{\rm c}$. Then, the following result characterizes, in terms of $\nu$, the contraction of $x_a$ and $x_b$ on $[t_0,t_1]$.
\begin{proposition}\label{prop.contraction-trajectories}
	The solutions $x_a$ and $x_b$ contract from $t_0$ to $t_1$ with respect to $d$ if and only if 
	\begin{equation*}
		\int_{T_{\rm c}} \nu(x_a(t),x_b(t))\,\mathrm{d}t >    - \int_{T_{\rm i}} \nu(x_a(t),x_b(t))\,\mathrm{d}t .
	\end{equation*}
\end{proposition}
Proposition \ref{prop.contraction-trajectories} is proved in Appendix~\ref{sec.proof.prop.contraction-trajectories}. It provides
a necessary and sufficient characterization for trajectory contraction in terms of the functional $\nu$.
A simpler, yet only sufficient criterion for exponential contraction of multiple trajectories is derived in the next proposition. This criterion only involves the fraction of time spent by the considered trajectories in the two contraction regions, instead of the full behavior of the trajectories as required by Proposition~\ref{prop.contraction-trajectories}.
Consider an input $u$, a set $X_0\subset\R^2$ of initial conditions, 
and two time instants $t_0,t_1\in\R$ such that $t_1\ge t_0$. Let $\cS(X_0,u,t_0)$ denote the set of solutions of \eqref{s.FN} corresponding to the input $u$ and starting in $X_0$ at time $t_0$. Pick $\mu>0$ arbitrarily and 
let $\Delta_{\rm c}(\mu)$ denote the total amount of time spent by all the considered  trajectories, together,  in  $C^\uparrow(\mu)$ or in $C_\downarrow(\mu)$. Then, the following holds.

\begin{proposition}\label{prop.contraction-dwell-time}
	With $\mu>0$, let $\rho(\mu)\coloneqq1+\min\{\mu,b\varepsilon\}$ and $\alpha(\mu) \coloneqq e^{-\rho(\mu) \Delta_{\rm c}(\mu)   + t_1-t_0}$.
	If 
	\begin{equation}\label{e.dwell-time-condition}
		\Delta_{\rm c}(\mu) >  \frac{t_1-t_0}{\rho(\mu)} 
	\end{equation}
	then $\alpha(\mu)\in[0,1)$ and
	every two solutions in $\cS(X_0,u,t_0)$ $\alpha(\mu)$-contract from $t_0$ to $t_1$ with respect to the metric \eqref{d.d}. 
\end{proposition}  
Proposition~\ref{prop.contraction-dwell-time} is proved in Appendix~\ref{sec.proof.prop.contraction-dwell-time}.
The condition is deliberately conservative, since it combines worst-case contraction and expansion rates and requires all trajectories to occupy the same contraction region simultaneously. Its failure therefore does not imply lack of contraction; it only means that contraction cannot be certified from residence times alone.

\begin{figure*}
	\centering
	\begin{subfigure}{0.32\textwidth}
		\includegraphics[width=\textwidth,clip,trim=.4cm 0 .7cm 0]{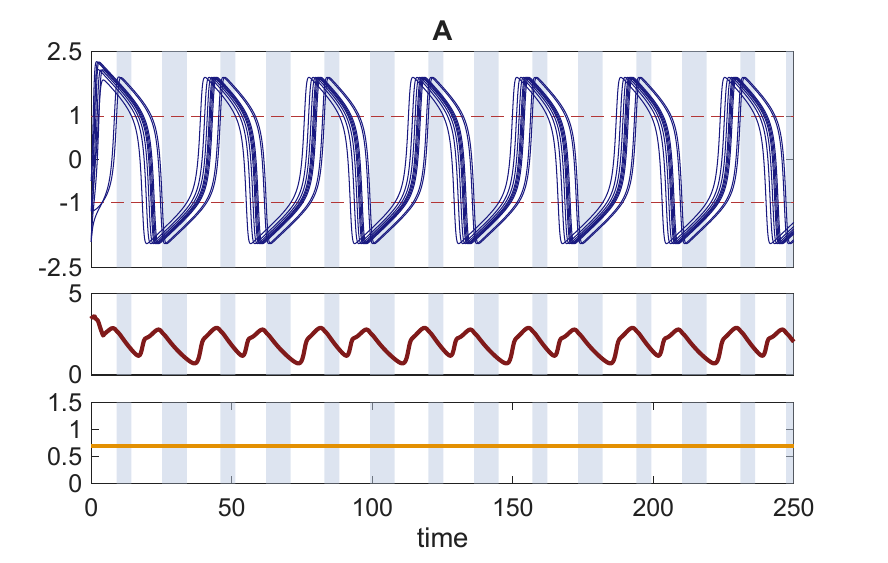} 
	\end{subfigure}
	\begin{subfigure}{0.32\textwidth}
		\includegraphics[width=\textwidth,clip,trim=.4cm 0 .7cm 0]{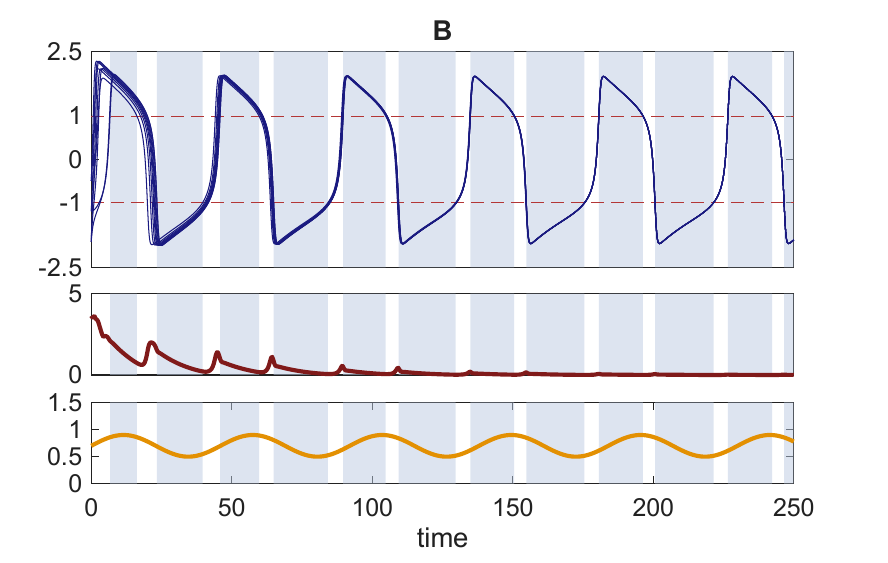} 
	\end{subfigure}
	\begin{subfigure}{0.32\textwidth}
		\includegraphics[width=\textwidth,clip,trim=.4cm 0 .7cm 0]{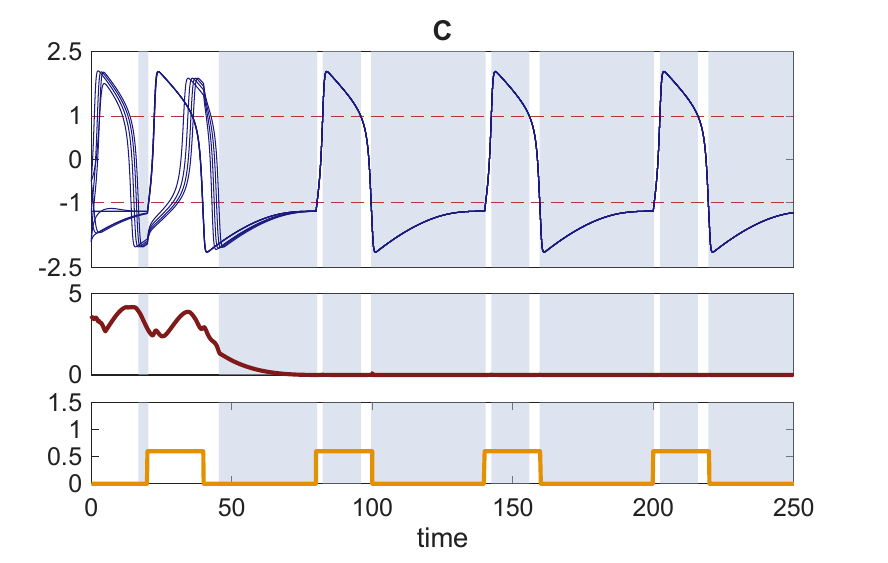} 
	\end{subfigure}
	
	\begin{subfigure}{0.32\textwidth}
		\includegraphics[width=\textwidth,clip,trim=.4cm 0 .7cm 0]{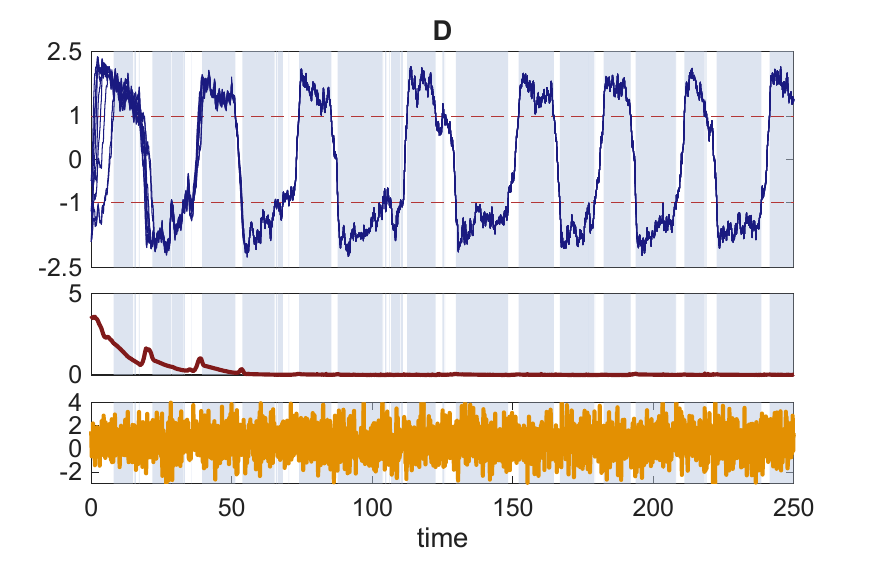} 
	\end{subfigure}
	\begin{subfigure}{0.32\textwidth}
		\includegraphics[width=\textwidth,clip,trim=.4cm 0 .7cm 0]{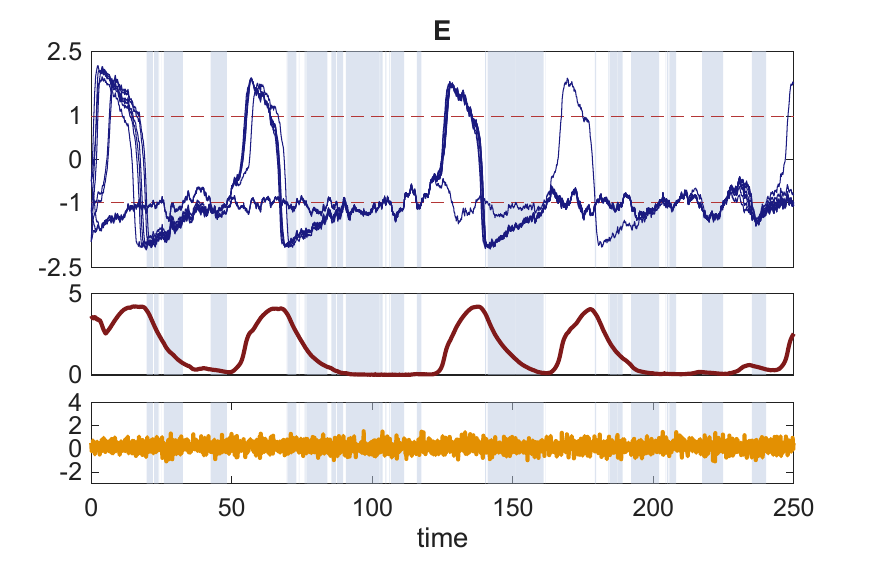} 
	\end{subfigure}
	\begin{subfigure}{0.32\textwidth}
		\includegraphics[width=\textwidth,clip,trim=.4cm 0 .7cm 0]{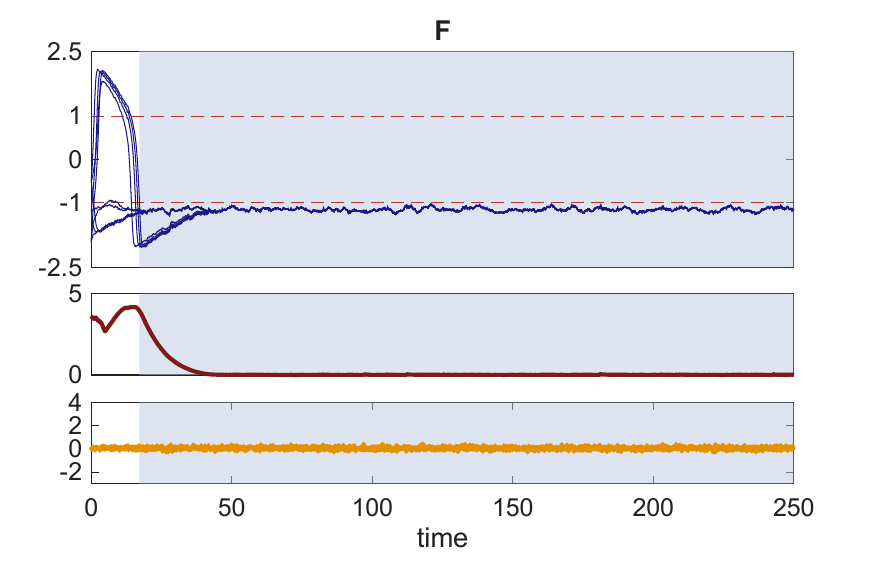}
	\end{subfigure}
	
	\caption{\textbf{(Un)reliability of the FitzHugh-Nagumo model.}
		All figures are obtained by simulating \eqref{s.FN} with $a=0.7$, $b=0.8$, and $\varepsilon=1/12.5$, and using \emph{MATLAB}'s \emph{ode23s} solver with output sampling resolution $10^{-3}$.
		Each figure depicts nine solutions corresponding to the initial conditions obtained by taking all   nine combinations of $v(0)\in\{-1.8994,\,-1.1994,-0.4994\}$ and $w(0)\in\{-1.3243,\,-0.6243,\,0.0757\}$.
		The noise signals are obtained, by means of a multiplication and addition of a bias value, from a common noise signal generated by linearly interpolating $2500$ samples of a Gaussian process distributed in a uniform grid covering the whole time interval $[0,250]$ and generated using \emph{MATLAB}'s \emph{randn} command with seed $123$. 
		In all figures: light-blue patches highlight the time intervals where all the considered solutions lie within the same contraction region; \textbf{(top)} the solid, dark-blue lines depict the time evolution of the $v$ variable for each of the nine solutions;  \textbf{(center)} ensemble diameter $\max_{a,b \in \{1,\ldots,9\}} d(x_a(t),x_b(t))$ under the metric \eqref{d.d}; \textbf{(bottom)} applied input $u$.
		Regarding the specific figures: \textbf{(A)} high constant input $u(t)=0.7$, contraction and expansion balance out resulting in phase-locking; \textbf{(B)} ``resonant input'' $u(t)= 0.7 + 0.2\sin(\pi t/23)$,  on average solutions contract; \textbf{(C)} slow periodic pulsating input (period $60$, amplitude $0.6$,  duty-cycle $1/3$),  solutions spend more time in the lower contraction region and, overall, they contract; \textbf{(D)} high Gaussian noise (standard deviation $1.2$, mean $0.6$),   solutions spend most of the time in the same contraction region and, on average, contract; \textbf{(E)} moderate Gaussian noise (standard deviation $0.4$, mean $0.2$),  solutions stay in the same contraction region very rarely, resulting in an unreliable behavior; \textbf{(F)} weak Gaussian noise (standard deviation $0.12$, bias $0.06$),    solutions are all eventually in the lower contraction region, they converge to the same subthreshold regime. }
	\label{fig.reliability}
\end{figure*}

\subsection{Numerical Illustration} \label{sec.reliability.contraction-realiability}
This section uses six representative  scenarios to illustrate qualitatively how the contraction analysis developed above accounts for reliability and unreliability across repeated trials. Each trial is represented by a solution driven by the same input but initialized at a different state. Reliability is assessed numerically from the convergence of this finite ensemble of trajectories; the contraction and expansion regions provide a geometric interpretation of the observed behavior.

The first scenario considers
a constant input $u(t)=0.7$. For the parameters considered here, this constant input induces an attractive limit cycle where the $v$ variable exhibits a sustained periodic oscillation (Fig. \ref{fig.reliability}-\textbf{A}). Trajectories starting from different initial conditions approach the same stable limit cycle but generally retain different asymptotic phases. The resulting behavior is therefore not reliable in the sense adopted here. The lack of trial-to-trial
reliability under constant-current stimulation  is well-documented experimentally%
\footnote{The experiments in \cite{mainen_reliability_1995} do not,
	however, display the persistent fixed phase offsets generated by the
	deterministic FHN model in Fig.~\ref{fig.reliability}-\textbf{A}.
	The apparent trial-to-trial phase drift may reflect intrinsic
	variability and additional neuronal dynamics not represented by
	\eqref{s.FN}. Nevertheless, \cite[Fig.~1A]{mainen_reliability_1995}
	shows approximately repetitive firing within individual trials.}%
\citep{mainen_reliability_1995,hunter_resonance_1998}.
From the
viewpoint of Section~\ref{sec.reliability.contraction}, this behavior is explained by the fact that the contraction accumulated along some portions of each cycle is asymptotically balanced by the expansion accumulated along the remaining portions, so that the integral of $\nu$ (Proposition~\ref{prop.contraction-trajectories}) over a complete cycle has no net contracting effect.  

\begin{figure}
	\centering
	\includegraphics[width=7cm,trim=.4cm 0cm .7cm 1cm]{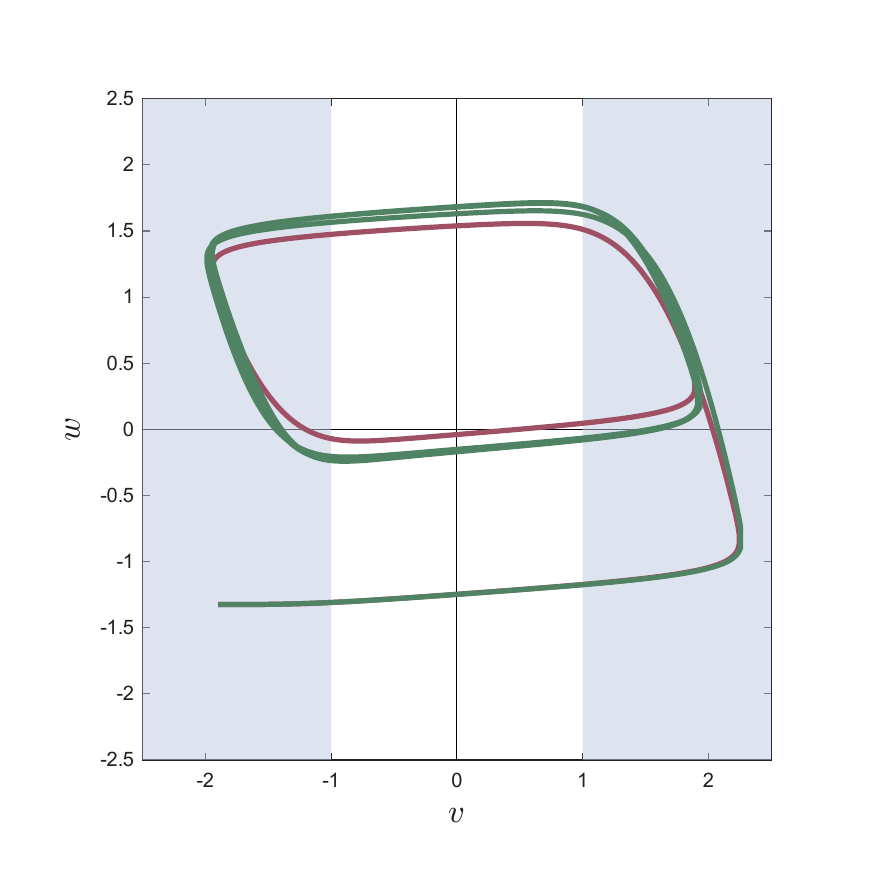} 
	
	\caption{\textbf{State-space visualization.} Plot in the $(v,w)$-plane of the trajectory originating in $(-1.8994,-1.3243)$ of the first \textbf{(red)} and the second \textbf{(green)} scenario, corresponding to Fig.~\ref{fig.reliability}-\textbf{A} and Fig.~\ref{fig.reliability}-\textbf{B}, respectively.}
	
	\label{fig.sspace}
\end{figure}

In the second scenario, a small harmonic component with a frequency close to the natural frequency of the limit-cycle oscillation observed in the first scenario is added to the constant stimulus. As shown in Fig.~\ref{fig.reliability}-\textbf{B}, trajectories starting from different initial conditions now converge toward the same response, resulting in a reliable behavior. This observation is consistent with the \emph{resonance effect} reported experimentally in~\cite{hunter_resonance_1998}, whereby periodic forcing near the neuron's natural firing frequency enhances spike-time reliability.
As shown in Fig.~\ref{fig.sspace}, trajectories draw a longer segment in the contraction regions compared to the first scenario, thereby suggesting an increase in the contraction balance over the interval.

The third scenario applies a pulsating input with period significantly \emph{larger} than the natural one. The observed behavior, depicted in Fig.~\ref{fig.reliability}-\textbf{C}, is reliable. Consistently with Proposition~\ref{prop.contraction-dwell-time}, reliability follows as a consequence of canonical dwell-time effects in switched systems~\citep{hespanha_stability_1999}  because the state spends most of the time close to its resting point in the lower contraction region, which is asymptotically stable with the parameters considered here. Using this principle, contraction of more general conductance-based models forced by slow pulsating input has also been proved in~\cite{cecconi_contraction_2026}.

The last three scenarios consider random inputs of different overall amplitudes, obtained by rescaling the same frozen-noise signal.
Fig.~\ref{fig.reliability}-\textbf{D} presents simulations with an input given by a linearly interpolated realization of a Gaussian process with mean $0.6$ and standard deviation $1.2$.
Consistent with the experimental evidence in~\cite{mainen_reliability_1995,hunter_resonance_1998}, the trajectories contract on average, thereby leading to a reliable behavior. The joint visitation pattern of the contraction regions in Fig.~\ref{fig.reliability}-\textbf{D} is similar to that observed under harmonic forcing in Fig.~\ref{fig.reliability}-\textbf{B}, and so is the resulting contraction dynamics. This observation is compatible with the conclusion of \cite{hunter_resonance_1998} that random inputs containing non-negligible power near the resonant frequency can enhance reliability through the resonance effect.

Fig.~\ref{fig.reliability}-\textbf{E} considers the same frozen-noise waveform scaled to a third its previous amplitude.
This input is sufficiently strong to drive the trajectories out of the lower contraction region, but not strong enough to bring them into the upper contraction region, except during a few isolated spikes. Hence, the trajectories spend most of the simulated interval in the expansion region, where contraction is not guaranteed. The corresponding behavior is not reliable. Unreliability is however  of a different flavor compared to Fig.~\ref{fig.reliability}-\textbf{A}, as in this case trajectories are mostly subthreshold. 

Finally, Fig.~\ref{fig.reliability}-\textbf{F} considers the same waveform scaled to one tenth of the amplitude used in Fig.~\ref{fig.reliability}-\textbf{D}. After a brief transient, this weak input no longer drives the simulated trajectories out of the lower contraction region, where they remain throughout the rest of the observation window.
As a result, all solutions contract exponentially after a brief transient, approaching a common subthreshold regime. This behavior is characteristic of convergence in the sense of \cite{pavlov_convergent_2004,pavlov_uniform_2006}, which is implied by contraction on a compact positively invariant set~\citep{ruffer_convergent_2013}.

Taken together, the last three scenarios display an   alternation between reliability and unreliability as the overall amplitude of the common input is varied. This alternation is qualitatively compatible with the findings of \cite{goldobin_antireliability_2006}: weak forcing produces reliability (Fig.~\ref{fig.reliability}-\textbf{F}), intermediate forcing produces unreliability (Fig.~\ref{fig.reliability}-\textbf{E}), and strong forcing restores reliability (Fig.~\ref{fig.reliability}-\textbf{D}). It should be noted, however, that the mechanism of reliability may be different from that  underlying the findings of \cite{goldobin_antireliability_2006}; for instance, in the last scenario of Fig.~\ref{fig.reliability}-\textbf{F} reliability may be attributed to asymptotic stability of the resting point. It should also be noted that the behavior observed in Figs.~\ref{fig.reliability}-\textbf{D--F} is not generic since, over a finite interval, the contraction behavior of the FHN model in the presence of noise is quite sensitive to the initial conditions and the specific input realization.

\section{Contraction of Linearly-Coupled FitzHugh-Nagumo Networks}\label{sec.networks}

The rationale behind Theorem~\ref{thm.contraction} extends rather naturally to study reliability of a class of neural networks composed of FHN models of the form \eqref{s.FN} interconnected through affine voltage couplings according to an arbitrary directed topology.
 
In particular, consider a network of $n\in\N_{\ge 1}$ FHN models, where the $i$th neuron is described by the following equations 
\begin{equation} \label{s.FNi} 
	\begin{aligned}
		\dot v_i &= v_i-\tfrac{1}{3}v_i^3 -w_i + \gamma_{i} u + \sum_{j=1}^n k_{ij} (v_j+c_j)  \\
		\dot w_i &= \varepsilon_i(v_i-b_i w_i+ a_i)   
	\end{aligned}
\end{equation}
in which $x_i\coloneqq(v_i,w_i)$ denotes the state of the $i$th system of the network, $u$ is the input to the network, affecting the $i$th system through the coefficient $\gamma_i\in\R$, $\varepsilon_i$, $b_i$, $a_i$ are the characteristic neuron coefficients, satisfying constraints analogous to \eqref{d.a_b_epsilon}, and, for each $j=1,\dots,n$,  $x_j\coloneqq(v_j,w_j)$ is the state of the $j$th FHN system,   $k_{ij}\in\R$ is a corresponding coupling gain, and $c_j\in\R$ represents a possible bias term. 
Within this phenomenological coupling description, a positive coefficient $k_{ij}$ represents an effective excitatory influence of neuron $j$ on neuron $i$, whereas negative $k_{ij}$ represents an effective inhibitory influence. 

Let $x\coloneqq (x_1,\ldots,x_n) \in \R^{2n}$ denote the aggregate network state and, with 
\begin{equation*}
	P\coloneqq \frac{1}{2}\diag\big(1,\varepsilon_1\inv,1,\varepsilon_2\inv,\ldots,1,\varepsilon_n\inv\big),
\end{equation*}
define the following metric on $\R^{2n}$
\begin{equation}\label{d.d-net}
	d(x,x')\coloneqq \sqrt{(x-x')\T P (x-x')}.
\end{equation}
This metric generalizes \eqref{d.d} to the network case.
Moreover, with 
\begin{equation*}
	K \coloneqq\begin{bmatrix}
		k_{11} & k_{12} & \cdots & k_{1n}\\
		k_{21} & k_{22} & \cdots & k_{2n}\\
		\vdots & \vdots & \ddots & \vdots \\
		k_{n1} & k_{n2} & \cdots & k_{nn} 
	\end{bmatrix}
\end{equation*}
denoting the \emph{coupling matrix}, define its symmetric part as
\begin{equation}
	K_{\mathrm{sym}}\coloneqq \frac{1}{2}(K +K\T).
\end{equation}
Then, the following theorem generalizes Theorem~\ref{thm.contraction} to the considered network setting.
\begin{theorem}\label{thm.contraction-net}
	Let $\mu\in(\R_{>0})^n$ be such that
	\begin{equation}\label{e.contraction-net.mu_g_K}
		\diag(\mu) - K_{\mathrm{sym}} \succ 0.
	\end{equation}
	Then, the network defined by the systems \eqref{s.FNi} for $i=1,\ldots,n$ is contractive on every convex subset of
	\begin{equation*}
		C(\mu)\coloneqq \{ x\in\R^{2n} \st \forall i=1,\ldots,n,\ v_i^2\ge 1+\mu_i \}
	\end{equation*}
	with respect to the metric \eqref{d.d-net}   with contraction
	rate
	\begin{equation*} 
			\lambda(\mu)
			\coloneqq
			\min\big\{ \lambda_{\min}(\diag(\mu)-K_{\mathrm{sym}}), 
			b_1\varepsilon_1,\ldots,b_n\varepsilon_n \big\}. 
	\end{equation*}
\end{theorem}

Theorem \ref{thm.contraction-net} is proved in Appendix~\ref{sec.proof.thm.contraction-net}.
It is worth noting that a sufficient condition for \eqref{e.contraction-net.mu_g_K} is that $\min_{i=1,\ldots,n} \mu_i > \lambda_{\max}(K_{\mathrm{sym}})$.
In the single-neuron case, $K_{\mathrm{sym}} = K = k_{11} = 0$; hence, Theorem~\ref{thm.contraction} is recovered. 
In the general case of $n$ neurons, the set $C(\mu)$ -- which plays the role of $C_{\downarrow}(\mu)\cup C^{\uparrow}(\mu)$ in the single-neuron case -- identifies $2^n$ contraction regions. These are separated by an expansion region $I\coloneqq\{ x\in\R^{2n}\st \exists i=1,\ldots,n,\ v_i^2\le 1\}$ where contraction may not hold. Hence, Theorem~\ref{thm.contraction-net} shows that the same geometric characterization of contraction given by Theorem~\ref{thm.contraction} for the single-neuron case is maintained in the general network case.  Accordingly, reliability of networks can be characterized in the same average-contraction terms discussed in Section~\ref{sec.reliability} for the single-neuron case.

\begin{figure}
	\centering 
	
	\begin{minipage}[t]{0.21\textwidth}
		\textbf{A} \\
		\includegraphics[width=3cm,clip,trim=0 0 0 0]{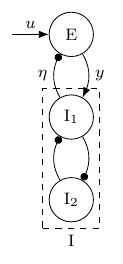}
	\end{minipage}
	\begin{minipage}[t]{0.21\textwidth}
		\textbf{B} \\
		\includegraphics[width=3cm,clip,trim=0 0 0 0]{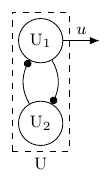} 
	\end{minipage}  
	
	\caption{\small
		\textbf{(A)} The EI network; \textbf{(B)} the exosystem. Links terminating with an arrow (bullet) denote excitatory (inhibitory) connections.
	} 
	\label{fig.EI}
\end{figure}

\section{Implications for Regulation} \label{sec.regulation}

As an application of the concepts and results of the previous sections, this section studies the reliability of a proof-of-concept example of a network showing how contraction can be seen as an enabler of network regulation. In this case, regulation leads to a homeostatic regime where a given output of the network becomes insensitive to a class of inputs. Then, this elementary motif is used to show how reliability and regulation can be used to devise a signal replicator network that is capable of propagating across an arbitrary finite number of stages a class of oscillatory inputs with zero phase lag. 

\subsection{The EI network}\label{sec.regulation.EI}

The motif depicted in Fig.~\ref{fig.EI}-\textbf{A},  which we will refer to as the \emph{EI network},  is a basic excitatory-inhibitory motif where neurons are modeled by FHN equations interconnected through affine couplings as in the previous Section~\ref{sec.networks}.
It was introduced in \cite{cecconi2025NOLCOS} as the solution of the following \emph{design} question. We are given a neuron -- $\rm E$ in Fig.~\ref{fig.EI}-\textbf{A} -- described by the following FHN model
\begin{equation}\label{s.E}
	{\rm E}\! : \!	\begin{syseq*}
		\dot v_{\e} &= v_{\e}-\tfrac{1}{3}v_{\e}^3 -w_{\e} + k_{\e}(u -\eta+V_{\rm rest}-y) \\
		\dot w_{\e} &= \varepsilon_{\e}(v_{\e}-b_{\e} w_{\e}+a_{\e})\\
		y&=v_{\e} 
	\end{syseq*}
\end{equation}
in which $a_\e$, $b_\e$, $\varepsilon_\e$ are the model parameters, subject to a constraint of the kind~\eqref{d.a_b_epsilon}, $V_{\rm rest}$ denotes the rest point of the variable $v_\e$ in the absence of coupling (\emph{i.e.}, with $k_\e=0$), and the term $k_{\e}(u -\eta+V_{\rm rest}-y)$, with $k_\e>0$, models a diffusive coupling with an 
excitatory input  $u$ representing an undesired input to be filtered out, and an inhibitory  input $\eta$ representing a ``control input'' we can design with the aim of counteracting $u$.  
Specifically, the design question is to construct, for a given class of inputs $u$, a neural network acting as a control system with the objective of making the output $y$ of $\mathrm{E}$ asymptotically insensitive to $u$. Namely, the network must reach homeostasis with respect to each $u$ in the considered class. 

In control-theoretic terms, this design question is an instance of the \emph{output regulation problem}, which is analytically solvable under the assumption that the disturbance $u$ is generated by a \emph{known} system, called the \emph{exosystem}~\citep{francis_internal_1975,bin_internal_2022}. 
In the illustrative context of this section, the exosystem is taken as a simple \emph{half-center oscillator}, which provides the most basic multi-neuron oscillatory motif~\citep{marder_central_2001}. 
This oscillator is modeled by system $\rm U$ in Fig.~\ref{fig.EI}-\textbf{B} as the reciprocal inhibitory  diffusive  interconnection of two FHN models of the form~\eqref{s.FN}. Its equations are
\begin{equation}\label{s.U}
	{\rm U}\! : \!	\begin{syseq*}
		\dot v_{\u1} &= v_{\u1}-\tfrac{1}{3}v_{\u1}^3 -w_{\u1} + k_{\u}(-v_{\u2}-v_{\u1})\\
		\dot w_{\u1} &= \varepsilon_{\u}(v_{\u1}-b_{\u} w_{\u1}+a_{\u})\\
		\dot v_{\u2} &= v_{\u2}-\tfrac{1}{3}v_{\u2}^3 -w_{\u2} + k_{\u}(-v_{\u1}-v_{\u2})\\
		\dot w_{\u2} &= \varepsilon_{\u}(v_{\u2}-b_{\u} w_{\u2}+a_{\u})\\
		u  &= v_{\u1},
	\end{syseq*}
\end{equation} 
in which $a_\u$, $b_\u$, $\varepsilon_\u$ are subject to a constraint of the kind~\eqref{d.a_b_epsilon} and $k_\u>0$.
It is worth noting that $\mathrm{U}$ models a whole class of inputs $u$ -- namely, all outputs of \eqref{s.U} generated by all admissible initial conditions -- and not only a single specific signal.

Under the assumption that $u$ is generated by the exosystem $\rm U$, the \emph{internal model principle}~\citep{bin_internal_2022} suggests that a control network may be designed by properly embedding an internal model of the exosystem. A minimal network with this ``internal model property'' is given by the block $\rm I$ in Fig.~\ref{fig.EI}-\textbf{A}, which is described by the  following equations
\begin{equation}\label{s.I}
	{\rm I}\!: \!	\begin{syseq*}
		\dot v_{\i1} &= v_{\i1}-\tfrac{1}{3}v_{\i1}^3 -w_{\i1} + k_{\u}(y-V_{\rm rest}-v_{\i2}-v_{\i1})\\
		\dot w_{\i1} &= \varepsilon_{\u}(v_{\i1}-b_{\u} w_{\i1}+a_{\u})\\
		\dot v_{\i2} &= v_{\i2}-\tfrac{1}{3}v_{\i2}^3 -w_{\i2} + k_{\u}(-v_{\i1}-v_{\i2})\\
		\dot w_{\i2} &= \varepsilon_{\u}(v_{\i2}-b_{\u} w_{\i2}+a_{\u}) \\
		\eta &= v_{\i1}.
	\end{syseq*}
\end{equation} 
System \eqref{s.I} is interconnected in feedback with system $\rm E$.
Yet, when $y=V_{\rm rest}$, system \eqref{s.I} becomes autonomous and equals the exosystem \eqref{s.U}. Hence, $\rm I$ embeds an internal model of $\rm U$.
As a direct consequence of such an internal model property,  for every disturbance $u$ generated by \eqref{s.U}, there exists an initial condition for the whole EI network \eqref{s.E}, \eqref{s.I} such that the corresponding solution satisfies $y(t)=V_{\rm rest}$ for all $t\in\R$ -- \emph{i.e.}, this solution is a homeostatic regime along which the undesired input $u$ is completely rejected. This initialization is obtained by taking the initial state of $\rm E$ equal to the resting point, and that of $\rm I$ matching the initialization of $\rm U$ producing $u$.

\begin{figure*}
	\centering

	\begin{subfigure}{.32\linewidth}
		\centering
		\small\textbf{A}\\
		\includegraphics[width=5cm,clip,trim=1cm .2cm 1.4cm 1.2cm]{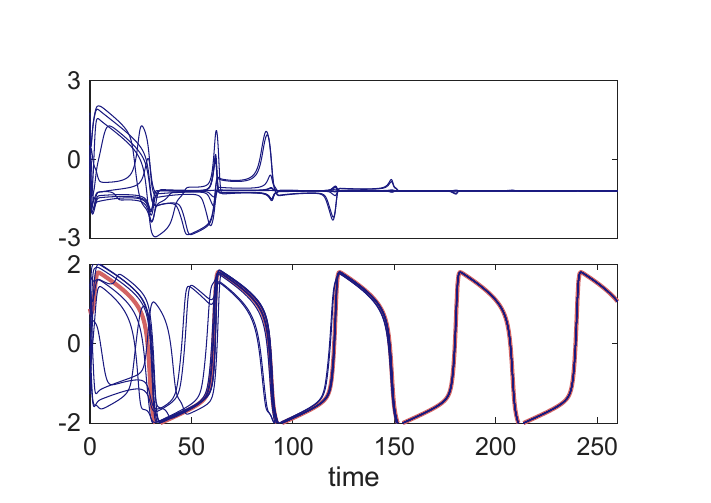} 
	\end{subfigure} 
	\begin{subfigure}{.32\linewidth}
		\centering
		\small\textbf{B}\\
		\includegraphics[width=5cm,clip,trim=1cm .2cm 1.4cm 1.2cm]{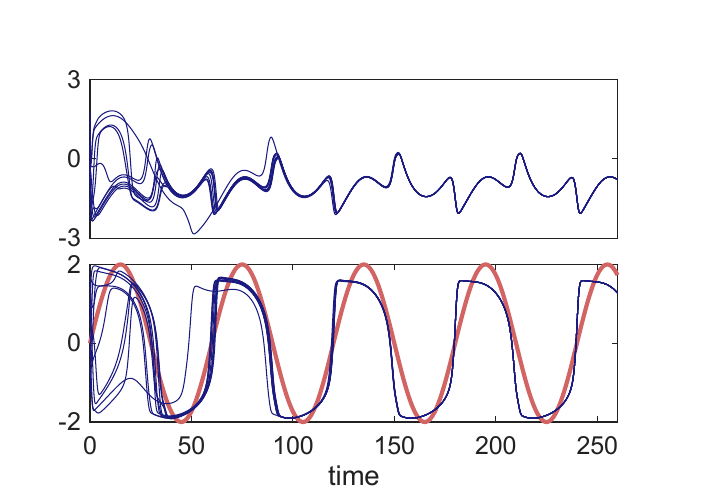} 
	\end{subfigure} 
	\begin{subfigure}{.32\linewidth}
		\centering
		\small\textbf{C}\\
		\includegraphics[width=5cm,clip,trim=1cm .2cm 1.4cm 1.2cm]{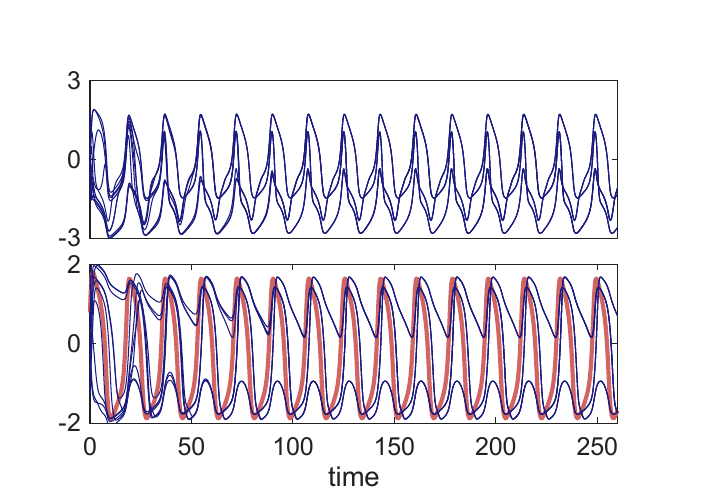} 
	\end{subfigure}
	
	\caption{\small
		Simulations of $10$ trials of the EI network \eqref{s.E}, \eqref{s.I} with $a_\e=0.7$, $b_\e=0.8$, $\varepsilon_\e=1/12.5$, $k_\e=4$, $a_\u=0.6$, $b_\u=0.7$,   $k_\u=1/2$, and $\varepsilon_\u=1/30$. Initial conditions are randomly chosen uniformly on $[-2,2]^6$ with seed $123$. \textbf{(A)} $u$ is generated by the exosystem \eqref{s.U}; \textbf{(B)} $u(t)= 2\sin(\pi t/30)$; \textbf{(C)} $u$ is generated by \eqref{s.U}   with  $\varepsilon_\u=1/5$ (while in \eqref{s.I}, $\varepsilon_\u$ remains $1/30$). 
		In all figures: \textbf{(top)} the $10$ simulated trajectories of $y(t)$; \textbf{(bottom)} dark blue lines depict the $10$ simulated trajectories of the variable $\eta$, red lines depict the disturbance $u$.    
	} 
	\label{fig.regulation}
\end{figure*}

\subsection{Reliability of the EI network}\label{sec.regulation.EI-reliability}

On such an ideal homeostatic regime, the effect of $y$ on the $\mathrm{I}$ system is null. Hence, $\mathrm{I}$ delivers a \emph{feedforward} action synchronized to the exosystem $\mathrm{U}$ and thus inhibiting the signal $u$ completely. 
For all other solutions, $y$ acts instead as a \emph{feedback} signal trying to properly re-phase $\mathrm{I}$ to bring the overall network toward the homeostatic regime.  
Asymptotic rejection of $u$ from $y$ is indeed guaranteed if the homeostatic regime attracts all the other trajectories, at least from a set of relevant initial conditions. In turn, this is possible if the EI network \eqref{s.E}, \eqref{s.I} is contractive: the solution of the regulation problem at hand thus becomes a question of  network reliability.

To study the reliability of the EI network we can apply Theorem~\ref{thm.contraction-net}. In the EI network case, the metric \eqref{d.d-net} is
\begin{equation}\label{d.d-EI}
	d(x,x') = \sqrt{(x-x')\T P (x-x')} ,
\end{equation}
with $P\coloneqq\frac{1}{2} \diag(1,\varepsilon_\e\inv,1,\varepsilon_\u\inv,1,\varepsilon_\u\inv)$, and the symmetric part of the coupling matrix equals
\begin{equation}\label{d.K}
	K_{\mathrm{sym}} = \begin{bmatrix}
		-k_\e & \frac{1}{2}(k_\u-k_\e) & 0\\
		\frac{1}{2}(k_\u-k_\e) & -k_\u & -k_\u\\
		0 & -k_\u & -k_\u 
	\end{bmatrix}.
\end{equation}
As a corollary of Theorem~\ref{thm.contraction-net}, we then obtain the following.
\begin{corollary}\label{cor.contraction-net}
	Let $\mu_\e,\mu_\i>0$ be such that
	\begin{equation}\label{e.network-margins}
		\diag(\mu_\e,\mu_\i,\mu_\i)-K_{\mathrm{sym}}\succ0.
	\end{equation}
	Then, the EI network \eqref{s.E}, \eqref{s.I} is contractive
	on every convex subset of
	\begin{align*}
		C(\mu)\coloneqq
		\big\{x\in\R^6\st {}&
		v_\e^2\ge1+\mu_\e,\ 
		v_{\i1}^2\ge1+\mu_\i,\\
		&v_{\i2}^2\ge1+\mu_\i
		\big\}
	\end{align*}
	with respect to the metric \eqref{d.d-EI} and with contraction
	rate
	\begin{equation*}
		\begin{split}
			\lambda(\mu)
			\coloneqq
			\min\big\{
			&\lambda_{\min}(\diag(\mu_\e,\mu_\i,\mu_\i)-K_{\mathrm{sym}}),\\&
			b_\e\varepsilon_\e,
			b_\u\varepsilon_\u
			\big\}.
		\end{split}
	\end{equation*}
\end{corollary}
 
As commented in Section~\ref{sec.networks},  a sufficient condition for \eqref{e.network-margins} is that $\mu_\e$ and $\mu_\i$ are both larger than the  maximum eigenvalue $\lambda_{\max}(K_{\mathrm{sym}})$ of $K_{\mathrm{sym}}$ (which is itself non-negative  since $K_{\mathrm{sym}}$ is symmetric and $r\coloneqq\begin{bmatrix}
	0 & 1 & -1
\end{bmatrix}$ satisfies $r K_{\mathrm{sym}}r\T=0$).
Moreover, if $k_\e=k_\u$ -- namely, when the network is \emph{gain-balanced} -- then $\lambda_{\max}(K_{\mathrm{sym}})=0$.  In this case, condition \eqref{e.network-margins} reduces to $\mu_\e>0$ and $\mu_\i>0$, similarly to Theorem~\ref{thm.contraction}.

In view of Corollary~\ref{cor.contraction-net}, we can  conclude that,  if the input $u$ makes the EI network's trajectories originating around the homeostatic regime  spend enough time in the contraction regions, then the EI network contracts and homeostasis occurs. Remarkably, the numerical evidence suggests that the signals $u$ generated by the exosystem \eqref{s.U} -- whose internal model is included in the EI network --   belong to this reliable input class. Moreover, an intuitive argument is as follows: for each $u$ produced by \eqref{s.U}, the corresponding homeostatic trajectory is such that the state of system $\mathrm{E}$ is at the resting point -- hence in its lower contraction region -- while the two neurons forming system $\mathrm{I}$ oscillate in counter-phase inducing a resonance-like effect on each other of the same kind of the one observed for single neurons in Fig.~\ref{fig.reliability}-\textbf{B}. Hence, overall, the EI network behaves reliably.

As a consequence of reliability,  the EI network achieves homeostasis with respect to the disturbances produced by the exosystem~\eqref{s.U}, as
Fig.~\ref{fig.regulation}-\textbf{A} illustrates:  in response to a disturbance $u$ generated by \eqref{s.U}, in all trials $y(t)$ converges to $V_{\rm rest}$ (top) and $\eta(t)$ synchronizes with $u(t)$ (bottom). 
In addition, Fig.~\ref{fig.regulation}-\textbf{B} reports  a simulation in which $u$ is a sinusoid  similar in frequency and amplitude to the disturbance of Fig.~\ref{fig.regulation}-\textbf{A}. Although in this case perfect rejection is not possible, since the internal model differs from the generator of a sinusoid,   contraction still takes place. This simulation illustrates the robustness of input-selective contraction with respect to perturbations of the input and, consequently, also illustrates how sinusoids can entrain neural circuits. Moreover, we also observe that, although not identical, $\eta$ and $u$ synchronize, and, consequently, the output trajectory $y$ does not spike. In the sense of super-threshold dynamics, the undesired input $u$ is therefore effectively filtered out as it induces no super-threshold spike in $y$.  

Lastly, Fig.~\ref{fig.regulation}-\textbf{C} shows a simulation with an exosystem that is modified by setting $\varepsilon_\u=1/5$ while keeping $\varepsilon_\u=1/30$ in~\eqref{s.I}. In this case, the exosystem is significantly different from the $\mathrm{I}$ subsystem, and the internal model property is too far from being satisfied. Contraction is no longer observed over the simulated interval, the trajectories of $\eta$ and $y$ do not synchronize across trials, and homeostasis does not take place: unreliability has destroyed functionality.
 
 \begin{figure*}
	
	\centering
	\begin{subfigure}[T]{0.47\textwidth} 
		\includegraphics[width=\textwidth,clip]{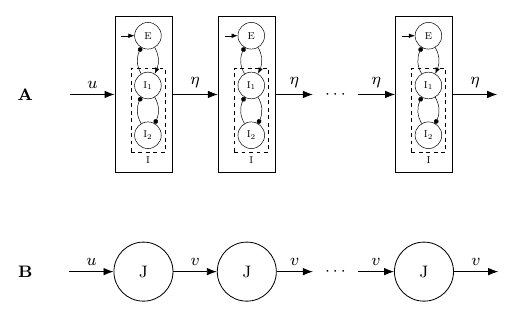} 
	\end{subfigure}
	\begin{subfigure}[T]{0.47\textwidth}
		\includegraphics[width=\textwidth,clip,trim=.4cm 0 .7cm .4cm]{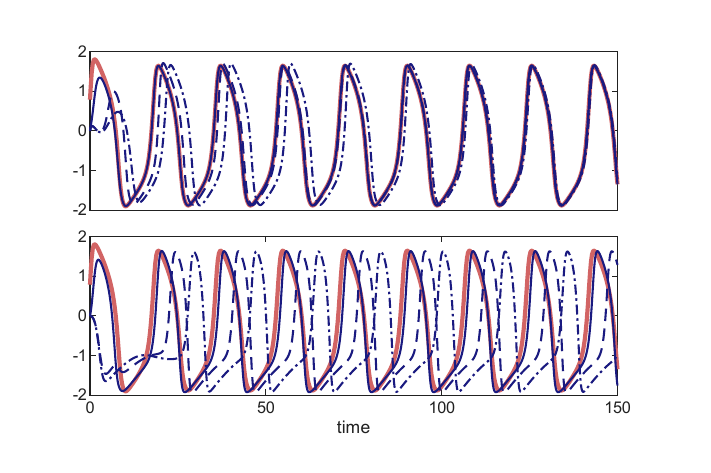}
	\end{subfigure} 
	
	\caption{\textbf{Reliability enables precise propagation.} In all simulations, the input signal $u$ is generated by $\rm U$ (Equations~\eqref{s.U}) with $\varepsilon_{\u}=1/5$, $a_\u=0.6$, $b_\u=0.7$, $k_\u=1/2$, and with initial conditions drawn at random within $[-2,2]^4$ with seed $123$; the systems in the chains are simulated from a zero initial condition.  The $\mathrm{E}$ subsystem of the EI network has the same parameters as in the experiments of Fig.~\ref{fig.regulation}. The parameters of the $\mathrm{I}$ and those of $\mathrm{J}$ equal those used for the exosystem $\mathrm{U}$.  \textbf{(Left)} block diagram representing the two simulated chains; the $\rm EI$ networks are described by \eqref{s.E}, \eqref{s.I} whereas $\rm J$ by \eqref{s.J}. \textbf{(Right)} simulation results for chain \textbf{A} (top) and \textbf{B} (bottom). In red, the input $u$; in blue, solid line the output ($\eta$ for chain \textbf{A} and $v$ for chain \textbf{B}) of the first chain stage; in blue, dashed line, the output of the 5th stage; in blue, dash-dotted line, the output of the 10th stage. While the $\rm EI$ chain \textbf{A} preserves synchrony across the chain, the other chain introduces significant phase shifts.}
	\label{fig.chains}
\end{figure*}

\subsection{Tracking and Propagation}\label{sec.regulation.propagation}

To compensate $u$, the variable $\eta$ must asymptotically match it. Hence, the EI network may also be interpreted as performing \emph{tracking} of $u$ in the variable $\eta$ with zero phase lag. In the simulations of Fig. \ref{fig.regulation}-\textbf{A} (bottom) we indeed observe perfect tracking.
This tracking motif may serve as a building block for several more complex functional modules. For instance, a chain of $\rm EI$ networks can be used to   propagate $u$ downstream without accumulating phase lag. 
A simulation illustrating the benefit, in terms of propagation precision, of a chain of $\rm EI$ networks over simple neuron-to-neuron chains is shown in Fig.~\ref{fig.chains}. Therein, two chains are compared when subject to an input $u$ generated by the exosystem $\rm U$.
Chain~\textbf{A} is given by the concatenation of several $\rm EI$ networks where the output $\eta$ of each network constitutes the input to the next one. Chain \textbf{B} is given by the series concatenation of single neuron  models.  To foster a favorable resonance effect, these neurons are taken as FHN models with parameters matching those of the exosystem; namely, they are described by the following equation 
\begin{equation}\label{s.J}
	{\rm J}:\ 	\begin{syseq*}
		\dot v  &= v -\tfrac{1}{3}v^3 -w  + k_{\u}(u-v) \\
		\dot w  &= \varepsilon_{\u}(v -b_{\u} w +a_{\u}). 
	\end{syseq*}
\end{equation} 

The two chains are represented in Fig.~\ref{fig.chains} (left), while a simulation of such chains with $10$ stages is presented in Fig.~\ref{fig.chains} (right). As the figure shows, in the asymptotic regime the chain of $\rm EI$ networks (chain \textbf{A}) does not introduce phase lags between $u$ and the output $\eta$ of the last stage. Indeed, after the transient, the output of the last stage perfectly tracks the input $u$; the only visible effect is a transient whose length grows with the chain stage. 
Chain \textbf{B}, instead, introduces a phase lag that grows with the chain length.
This behavior can be formally established starting from the contraction of the EI network to the signals $u$ generated by the exosystem $\mathrm{U}$ either by a direct application of Theorem~\ref{thm.contraction-net} or by an induction argument, since  each stage is driven by an input (the output $\eta$ of the preceding stage) asymptotically matching $u$.

There is substantial experimental evidence that neuronal populations can become precisely entrained to rhythmic sensory stimulation.
For instance, \cite{rager_response_1998} found that cortical responses in cats exhibited precise phase locking to flickering visual stimuli over a broad range of stimulation frequencies (2–50 Hz). They hypothesize that
\emquote{there are actually strongly damped oscillators along the retino-cortical path which are tuned to different frequency bands and get recruited to variable extents at different stimulation frequencies} and that  \emquote{one possibility is that these internal synchronizing mechanisms contributed to the regularization of the steady state responses, thereby generating such perfect phase locking. This interpretation is supported by the finding that it took  several flash cycles before the responses got entrained into a regular oscillatory pattern}~\citep{rager_response_1998}.
The presence of the $\rm I$ subnetwork in each stage of the chain -- a neural oscillator --  is compatible with such findings. We also observe in Fig.~\ref{fig.regulation} and~\ref{fig.chains} the mentioned multi-cycle transient.

Stimulus-dependent synchronization
between spatially separated neuronal populations has also been observed
within cat visual cortex~\citep{engel_stimulusdependent_1990}. Such internally
coordinated synchrony has been proposed as a mechanism for dynamically
organizing distributed cortical responses~\citep{singer_neuronal_1999}.
Moreover, precise synchronization between neuronal populations in distant brain regions has also been reported experimentally~\citep{roelfsema_visuomotor_1997,brecht_correlation_1998,rodriguez_perceptions_1999}, and several explanatory models have been proposed~\citep{ritz_synchronous_1997,vicente_dynamical_2008,uhlhaas_neural_2009}. 
These include: feedforward \emph{pacemaker} motifs where autonomous oscillatory cells -- such as the \emph{chattering cells} of \cite{gray_chattering_1996} -- entrain other networks~\citep{ritz_synchronous_1997,uhlhaas_neural_2009}; propagation via \emph{synfire chains} \citep{abeles_corticonics_1991,aertsen_propagation_1996} -- although these typically introduce stage-dependent propagation delays, qualitatively resembling the phase-lag accumulation observed in chain \textbf{B} in Fig.~\ref{fig.chains}; chains formed by locally inhibitory subnetworks excited so as to produce spike doublets~\citep{whittington_synchronized_1995,traub_mechanism_1996,kopell_gamma_2000};
 \emph{dynamical relaying} motifs where an intermediary network fosters synchrony between two separated subnetworks  \citep{vicente_dynamical_2008,gollo_mechanisms_2014}.
 
 The $\rm EI$-chain  motif suggests another complementary proof-of-concept mechanism of precise propagation. This mechanism is compatible with the others in several respects. For instance, feedforward entrainment, as described in \citep{gray_chattering_1996,ritz_synchronous_1997,uhlhaas_neural_2009}, is indeed what happens in the homeostatic regime where feedback vanishes.
 The main difference is that, in the EI network, such a homeostatic regime is only one of the possible behaviors of an otherwise feedback system. In this connection, we stress that, while feedback and feedforward are often seen as alternative motifs~\citep{ritz_synchronous_1997}, they are instead unified in the $\rm EI$ chain model.
 Moreover, as in synfire chains, propagation is unidirectional. As in \citep{whittington_synchronized_1995,traub_mechanism_1996}, each $\rm EI$ stage embeds a subnetwork made of inhibitory connections that generates the rhythmic pattern. Finally, as in the dynamical relay approach of \citep{vicente_dynamical_2008,gollo_mechanisms_2014}, synchronization is due to an overall network effect. Here, the $\rm E$ systems of each two successive stages of an $\rm EI$ chain share a connection with a common $\rm I$ subnetwork, which therefore acts as an intermediary subnetwork.
 
 The EI cascade should therefore be regarded as a complementary control-theoretic proof of concept: internal-model dynamics generate the target regime where $u$ is replicated at the end of the chain, while reliability ensures asymptotic stability of such a regime. Whether the same mechanism can preserve near-zero phase relationships in the presence of synaptic couplings and heterogeneous synaptic and conduction delays remains an open question.

\section{Discussion}\label{sec.discussion}

The article has introduced a new state-space characterization of reliability based on contraction analysis for (networks of) FitzHugh-Nagumo models. It has been shown that the state space of a network of an arbitrary finite number $n$ of FHN systems, interconnected through affine couplings, contains $2^n$ contraction regions separated by an expansion region where contraction may not hold. In any given interval of time, an input $u$ of the network will lead to an overall contractive behavior if, within such an interval, the state of the network spends enough time in the contraction regions. This suggests that contraction, hence reliability, is inherently a matter of time scale: a behavior can be reliable on some interval, unreliable on a larger interval, and then again reliable on an even larger one.

The single-neuron analysis also shows that spiking can make a substantial contribution to contraction, since each full spike brings the state into the upper contraction region. Reliability therefore cannot generally be attributed solely to the stability of the resting state. Simulations done in \cite{cecconi2025NOLCOS} for a choice of the parameters leading to an unstable resting point further show that contraction can still be achieved, for instance with a resonant input of the kind used in scenario 2 of Section~\ref{sec.reliability.contraction-realiability}, if the resting point is unstable.  
Stability of the resting point plays instead a predominant role for slow inputs such as the third scenario of Section~\ref{sec.reliability.contraction-realiability}, or for weak inputs such as in the last scenario.

The EI network example developed in Section~\ref{sec.regulation} suggests that characterizing reliability in terms of contraction may open a new connection between reliability, the internal model principle, and regulation of neuronal networks. The input-dependent nature of reliability makes regulation an \emph{input-selective} property: it holds for some inputs, it does not for others. For inputs leading to a reliable behavior, the network becomes functional, as it consistently regulates. 
 For inputs not inducing a reliable behavior, the corresponding response is sensitive to the initial operating point. In this case, the response is not reproducible across repeated presentations of the same input. At the scale of multiple presentations of the same input, this irreproducibility may be interpreted as lack of functionality.  
 
 The mathematical setting considered here is deliberately minimal and has limited biological plausibility. The considered models neglect synaptic dynamics, transmission delays, intrinsic noise, and several forms of neuronal heterogeneity. As a consequence, the drawn conclusions remain speculative, and should be regarded as a proof of principle. Extending the analysis to conductance-based neuronal and synaptic models is an important direction for future work.
 
 Finally, from a control-theoretic viewpoint this interplay between resonance, input-selective contraction, and regulation establishes an interesting connection to classical regulation theory for \emph{linear time-invariant systems (LTI)}. In linear regulation, internal models are linear undamped oscillators (or integrators in the case of constant signals) driven by the regulation errors.
 Linear oscillators are resonators: if driven by a sinusoidal forcing at their natural frequency, they respond with an oscillation whose amplitude grows linearly with time, and whose phase only depends on the forcing input. As a consequence, the bounded homogeneous component depending on the initial conditions becomes asymptotically negligible relative to the growing forced component: the peak times of the responses generated with the same resonant input and with different initial conditions asymptotically align. This asymptotic alignment is a weaker form of input-selective reliability and, indeed, resonance is the basic mechanism at the core of linear regulation as it is in the EI network example.

\begin{appendices}

\section{Proof of Theorem~\ref{thm.contraction}}\label{sec.proof.thm.contraction}

By defining $x \coloneq(v, w)$, $f(x)\coloneq (v  - \frac{1}{3}v^3 - w   ,\, \varepsilon(v - bw + a) )$, and $B=\begin{bmatrix}
	1 & 0
\end{bmatrix}\T$,  we can rewrite \eqref{s.FN} in compact form as 
\begin{equation}\label{proof.s.x}
	\dot x = f(x) + Bu.
\end{equation} 
The Jacobian of $f(x)$  equals
\begin{equation*}
	 J(x) = \begin{bmatrix}
	 	1 - v^2 & - 1 \\ \varepsilon & - b \varepsilon
	 \end{bmatrix}.
\end{equation*}  
Moreover, by introducing the matrix $P = \frac{1}{2}\text{diag}(1, 1/\varepsilon)$, we deduce from \eqref{d.d} that, for every $x_a,x_b\in\R^2$, $d(x_a, x_b)^2 = (x_a - x_b)^{\top}P(x_a - x_b)$. Then, for every $\mu > 0$ and every $x$ in $C^{\uparrow}(\mu)$ or in $C_{\downarrow}(\mu)$, it holds that
\begin{equation}\label{proof.e.lyapunov_LMI}
	\begin{aligned}
		PJ(x) + J(x)^{\top}P &= \begin{bmatrix}
			1 - v^2 & 0 \\ 0 & - b
		\end{bmatrix} \preceq  \begin{bmatrix}
			-\mu & 0 \\ 0 & - b
		\end{bmatrix}.
	\end{aligned}
\end{equation}
Since $f$ is continuously differentiable, \emph{Hadamard's Lemma} implies 
\begin{equation}\label{proof.e.Hadamard}
	\begin{split}
		&f(x_a) - f(x_b) \\
		&\   =   \left( \int_0^1 J(sx_a + (1-s)x_b)\,\mathrm{d}s\right) (x_a - x_b)
	\end{split}
\end{equation} 
for all $x_a, x_b \in \R^2$.

Now, let $x_a, x_b$ be two solutions of \eqref{proof.s.x}, and consider any time instant $t$ such that both $x_a(t)$ and $x_b(t)$ lie in $C^{\uparrow}(\mu)$. By convexity of $C^{\uparrow}(\mu)$, we have $sx_a(t) + (1-s)x_b(t) \in C^{\uparrow}(\mu)$ for all $s \in [0,1]$. %
Define the variational variable $\tilde x(t) \coloneqq x_a(t) - x_b(t)$. Then,  \eqref{proof.e.lyapunov_LMI} and \eqref{proof.e.Hadamard} yield 
	\begin{align}
		\frac{\mathrm{d}}{\mathrm{d}t}&d(x_a(t), x_b(t))^2 
		= \frac{d}{dt}\left(\tilde x(t)^{\top}P\tilde x(t)\right) \nonumber\\ 
		&= \tilde x(t)^{\top}\bigg(\int_0^1 \Big(J\big(sx_a(t) + (1-s)x_b(t)\big)^{\top}P  \nonumber\\
		&\hspace{1.5cm}  + PJ\big(sx_a(t) + (1-s)x_b(t)\big)\Big)ds\bigg)\tilde x(t) \nonumber\\
		&\le \tilde x(t)^{\top}\begin{bmatrix} -\mu & 0 \\ 0 & -b \end{bmatrix}\tilde x(t)  = -\mu \tilde v ^2 - b \tilde w ^2 \nonumber\\
		&   \le -2\min\{ \mu , b\varepsilon\}\, d(x_a(t),x_b(t))^2 \label{proof.e.decay_d_in_contraction_regions}
	\end{align} 
where $\tilde v$ and $\tilde w$ are such that $\tilde x= (\tilde v, \tilde w)$.  
By \emph{Gr\"{o}nwall's Lemma}, it then follows that, if $x_a$ and $x_b$ stay within $C^{\uparrow}(\mu)$ for all times within $[t_0,t)$, then
\begin{equation*}
	d(x_a(t), x_b(t))^2 \le e^{-2\min\{\mu, b\varepsilon\}(t - t_0)}d(x_a(t_0), x_b(t_0))^2.
\end{equation*}
Taking the square root yields the result, and 
a similar argument holds for $C_{\downarrow}(\mu)$.

\section{Proof of Proposition~\ref{prop.contraction-trajectories}}\label{sec.proof.prop.contraction-trajectories}

Pick 
any two solutions $x_a=(v_a,w_a)$ and $x_b=(v_b,w_b)$ of \eqref{s.FN} subject to the same input $u$, and any two time instants $t_0,t_1\in\R$ such that $t_0< t_1$. Set $\tilde v\coloneqq v_a-v_b$ and $\tilde w\coloneqq w_a-w_b$. 
Since $v_a^3-v_b^3=  \tilde v (v_a^2+v_av_b+v_b^2)$, from \eqref{s.FN} one obtains
 \begin{align*}
  	\dot{\tilde{v}} &= \left(1 - \frac{1}{3} (v_a^2+v_av_b+v_b^2)\right)\tilde v  - \tilde w, & \dot{\tilde{w}} &= \varepsilon(\tilde v-b\tilde w)
 \end{align*}  
Since, for every $t\in\R$, $d(x_a(t),x_b(t))^2 = \frac{1}{2}\tilde v(t)^2 + \frac{1}{2\varepsilon} \tilde w(t)^2$, differentiating and using \eqref{d.nu} yields  
\begin{equation}\label{proof.e.d-sq_nu} 
	\begin{split}
		\frac{\mathrm{d}}{\mathrm{d}t}d(x_a(t),x_b(t))^2
		&= \left( 1-\frac{1}{3}(v_a^2+v_av_b+v_b^2)\right)\tilde v^2 - b\tilde w^2\\
		&= -\nu(x_a(t),x_b(t)) 
	\end{split} 
\end{equation}
for all $t\in[t_0,t_1]$. Integrating  from $t_0$ to $t_1$ yields that $d(x_a(t_1),x_b(t_1))<d(x_a(t_0),x_b(t_0))$ is equivalent to
\begin{equation*} 
	  \int_{t_0}^{t_1}\nu(x_a(t),x_b(t))\,\mathrm{d}t  > 0
\end{equation*}
Since $T_{\mathrm{i}}$ and $T_{\mathrm{c}}$ form a disjoint partition of $[t_0,t_1]$ the result follows.

\section{Proof of Proposition~\ref{prop.contraction-dwell-time}}\label{sec.proof.prop.contraction-dwell-time}
 
	Let $T_{\rm c}(\mu)$ denote the set of time instants in which all solutions in $\cS(X_0,u,t_0)$ are   either all in $C^\uparrow(\mu)$ or all in $C_\downarrow(\mu)$. Then, $\Delta_{\rm c}(\mu)$ equals the Lebesgue measure of $T_{\rm c}(\mu)$.
	Pick any two solutions $x_a,x_b\in\cS(X_0,u,t_0)$.
	Using \eqref{proof.e.decay_d_in_contraction_regions}, one obtains
	\begin{equation}\label{proof.e.decay-in-Tc}
	\frac{\mathrm{d}}{\mathrm{d}t} d(x_a(t),x_b(t))^2 \le -2(\rho(\mu)-1) d(x_a(t),x_b(t))^2
	\end{equation} 
	for all $t\in T_{\rm c}(\mu)$, where $\rho(\mu)\coloneqq1+\min\{\mu,b\varepsilon\}$.  
	Moreover, using \eqref{proof.e.d-sq_nu}, and noting that $v_a^2+v_b^2+v_av_b \ge0$ for all $v_a,v_b\in\R$, one obtains 
	\begin{equation}\label{proof.e.growth-outside-Tc}
		\frac{\mathrm{d}}{\mathrm{d}t} d(x_a(t),x_b(t))^2\le (v_a-v_b)^2 \le 2  d(x_a(t),x_b(t))^2 
	\end{equation}
	for all $t\in [t_0,t_1]\setminus T_{\rm c}(\mu)$.
	
	With $\chi_{\mathrm{c}}(t)$ the indicator function of $T_{\rm c}(\mu)$, 
	joining \eqref{proof.e.decay-in-Tc}   and \eqref{proof.e.growth-outside-Tc}  gives
	\begin{equation*}
		\begin{split}
			&\frac{\mathrm{d}}{\mathrm{d}t} d(x_a(t),x_b(t))^2
			\\
			&\quad \le \left( \chi_{\mathrm{c}}(t) 2 (1-\rho(\mu)) + (1-\chi_{\mathrm{c}}(t)) 2 \right) d(x_a(t),x_b(t))^2
			\\&\quad = 2\big(1-\rho(\mu)\chi_{\mathrm{c}}(t)\big) d(x_a(t),x_b(t))^2
		\end{split} 
	\end{equation*}
	for all $t\in[t_0,t_1)$. By \emph{Gr\"{o}nwall's Lemma}, we then obtain  
 \begin{equation*}
 	d(x_a(t_1),x_b(t_1)) \le\alpha(\mu) d(x_a(t_0),x_b(t_0)).
 \end{equation*} 
  If \eqref{e.dwell-time-condition} holds, then $\alpha(\mu)<1$, and the result follows.

\section{Proof of Theorem~\ref{thm.contraction-net}}\label{sec.proof.thm.contraction-net}

Define $v\coloneqq(v_1,\ldots,v_n)$, $w\coloneqq(w_1,\ldots,w_n)$, $E \coloneqq \diag(\varepsilon_1,\ldots,\varepsilon_n)$, $B\coloneqq\diag(b_1,\ldots,b_n)$,  and
\begin{align*}
	c &\coloneq \begin{bmatrix}
		\sum_{j=1}^n k_{1j}c_j \\ \vdots \\ \sum_{j=1}^n k_{nj}c_j 
	\end{bmatrix}  &      \Gamma&\coloneq \begin{bmatrix}
		\gamma_1 \\ \vdots \\ \gamma_n 
	\end{bmatrix}
	&  \\
	f(v)&\coloneqq  \begin{bmatrix}
		v_1-\frac{1}{3}v_1^3\\\vdots\\v_n-\frac{1}{3}v_n^3
	\end{bmatrix} &   a&\coloneqq\begin{bmatrix}
	a_1 \\ \vdots \\ a_n
	\end{bmatrix} .
\end{align*}
Then, the overall network, given by equations \eqref{s.FNi} for $i=1,\ldots,n$, can be compactly rewritten as
\begin{equation}\label{proof.s.vw}
	\begin{aligned}
		\dot v &=   f(v) - w + Kv + c + \Gamma u\\
		\dot w &= E(v-Bw+a) .
	\end{aligned}
\end{equation}
Denote $z\coloneqq(v,w)$.
Then, $z=Tx$ for some suitable permutation matrix $T\in\R^{2n\x 2n}$ and where, we recall, $x=(x_1,\ldots,x_n)$ with $x_i=(v_i,w_i)$ for all $i=1,\dots,n$.
The Jacobian of the vector field $z \mapsto  (f(v) - w + Kv + c +\Gamma u, \ E(v-Bw+a))$ describing the dynamics of \eqref{proof.s.vw}  is
\begin{equation*}
	J(z) = \begin{bmatrix}
		A(z) & -I\\
		E & -EB
	\end{bmatrix}
\end{equation*}
in which 
\begin{equation}\label{proof.d.A}
	A(z)\coloneq \diag(1-v_1^2,\ldots,1-v_n^2) + K.
\end{equation}
Define $Q\coloneqq\frac{1}{2}\diag(I,E\inv)$.  Then,
\begin{equation*}
		J(z)\T Q+Q J(z)  = \begin{bmatrix}
			 \tfrac{1}{2}(A(z)\T+A(z))& 0\\0 & -B  
		\end{bmatrix} .
\end{equation*} 
Let $\mu_1,\ldots,\mu_n>0$ be such that \eqref{e.contraction-net.mu_g_K} holds.
Then, in view of \eqref{proof.d.A}, for every $x\in C(\mu)$, $z=Tx$ satisfies
\begin{equation*}
	\begin{split}
		\tfrac{1}{2}(A(z)\T+A(z)) &= \diag(1-v_1^2,\ldots, 1-v_n^2) + K_{\mathrm{sym}}\\
		&\preceq - \diag(\mu_1,\ldots,\mu_n) + K_{\mathrm{sym}} \prec 0.
	\end{split}
\end{equation*} 
This implies
\begin{equation*} 
	 J(z)\T Q+Q J(z)  \preceq
	 	- 2\lambda(\mu) Q 
\end{equation*}
with $\lambda(\mu)=\min\{\lambda_{\min}(\diag(\mu)-K_{\mathrm{sym}}),b_1\varepsilon_1,\ldots,b_n\varepsilon_n\}$.

The claim then follows by means of the same arguments used in the proof of Theorem~\ref{thm.contraction} and by noticing that, for every $x,x'\in\R^{2n}$, one has that $z\coloneqq Tx$ and $z'\coloneqq Tx'$ satisfy $d(x,x')^2 =  (x-x')\T P (x-x')  = (z-z')\T (T\inv)\T P T\inv (z-z') = (z-z')\T Q(z-z')$.

\end{appendices}

\bibliography{biblio}%

\end{document}